\documentclass{aa}
\bibpunct{(}{)}{;}{a}{}{,} % to follow the A&A style ...
\usepackage{graphicx}
\usepackage{txfonts}
\usepackage{footnote}

\begin{document} 

\DeclareGraphicsExtensions{.pdf,.png}

\title{The Solar Twin Planet Search}
\subtitle{II. A Jupiter twin around a solar twin}

\author{M. Bedell\inst{1}\thanks{Visiting Researcher at the Departamento de Astronomia do IAG/USP, Universidade de S\~{a}o Paulo} \and
J. Mel\'{e}ndez\inst{2} \and
J. L. Bean\inst{1} \and
I. Ram\'{i}rez\inst{3} \and
M. Asplund\inst{4} \and
A. Alves-Brito\inst{5} \and
L. Casagrande\inst{4} \and
S. Dreizler\inst{6} \and
T. Monroe\inst{2} \and
L. Spina\inst{2} \and
M. Tucci Maia\inst{2}}

\institute{Department of Astronomy and Astrophysics, University of Chicago, 5640 S. Ellis Ave, Chicago, IL 60637, USA\\
\email{mbedell@oddjob.uchicago.edu}
\and
Departamento de Astronomia do IAG/USP, Universidade de S\~{a}o Paulo, Rua do Mat\~{a}o 1226, Cidade Universit\'{a}ria, 05508-900 S\~{a}o Paulo, SP, Brazil
\and
McDonald Observatory and Department of Astronomy, University of Texas at Austin, USA
\and
Research School of Astronomy and Astrophysics, The Australian National University, Cotter Road, Weston, ACT 2611, Australia
\and
Instituto de Fisica, Universidade Federal do Rio Grande do Sul, Av. Bento Goncalves 9500, Porto Alegre, RS, Brazil
\and
Institut f{\"u}r Astrophysik, University of G{\"o}ttingen, Germany
}


\abstract{With high-precision radial velocity surveys reaching a sufficiently long time baseline, the domain of long-period planet detections has recently opened up. The search for Jupiter-like planets is especially important if we wish to investigate the prevalence of solar system analogs, but their detection is complicated by the existence of stellar activity cycles on similar timescales.  Radial velocity data with sufficiently long-term instrumental precision and robust methods of diagnosing activity are crucial to the detection of extrasolar Jupiters.}{Through our HARPS survey for planets around solar twin stars, we have identified a promising Jupiter twin candidate around the star HIP11915. We characterize this Keplerian signal and investigate its potential origins in stellar activity.}{We carry out a Markov chain Monte Carlo (MCMC) analysis of the radial velocity data.  To examine the signal's origin, we employ a variety of statistical tests using activity diagnostics such as the Ca II H\&K lines and line asymmetry tracers.}{Our analysis indicates that HIP11915 hosts a Jupiter-mass planet with a 3800-day orbital period and low eccentricity.  Although we cannot definitively rule out an activity cycle interpretation, we find that a planet interpretation is more likely based on a joint analysis of RV and activity index data.}{The challenges of long-period radial velocity signals addressed in this paper are critical for the ongoing discovery of Jupiter-like exoplanets. If planetary in nature, the signal investigated here represents a very close analog to the solar system in terms of both Sun-like host star and Jupiter-like planet.}

\authorrunning{Bedell et al.}

\maketitle

\section{Introduction}
As the search for new exoplanets with the radial velocity (RV) technique continues, our detection capabilities have increased dramatically in two dimensions: smaller signals, such as those induced by short-orbit terrestrial-mass planets, can now be found thanks to the increased precision of modern RV spectrographs, and longer-period signals can also be identified thanks to an ever-increasing time baseline of data collection.  Both of these detection capabilities are critical if we wish to find planetary systems comparable to our own solar system.  

At the cutting edge of long-period planet detection is the search for Jupiter analogs.  In terms of signal size, Jupiter induces by far the largest radial velocity shift on the Sun of any planet and therefore should be the first planet detected in a twin of the solar system.  However, the need for more than a decade of observational coverage with a consistently precise RV zero point complicates the process of discovering such a planet, as does the possibility of stellar magnetic activity cycles inducing RV variations on a similar timescale \citep{Saar1997}.

In spite of these challenges, a number of Jupiter analog candidates have been discovered from long-duration planet surveys \citep[see, e.g.,][]{Vogt2000, Naef2003, Wright2007, Wright2008, Howard2010, Boisse2012, Marmier2013, Wittenmyer2014, Feng2015}.  These surveys also demonstrate the prevalence of solar-like magnetic activity cycles, which can manifest in the stellar radial velocities \citep{Lovis2011,Zechmeister2013}.  These cycles can generally be traced by their influence on activity diagnostics such as the Ca II H and K lines \citep{Baliunas1995} and asymmetries of the spectral cross-correlation function \citep{Dravins1985}.  With careful attention to these diagnostics, we can minimize the influence of activity cycles and recover planet signatures around cycling stars \citep{Dumusque2011}.

Over the last four years, we have been conducting an RV planet search with the HARPS spectrograph \citep{Mayor2003} targeting a specific subset of 63 stars selected by the proximity of their fundamental parameters to those of the Sun ($T_{\textrm{eff}}$ within $\sim$100 K, log(g) within $\sim$0.1 dex, and [Fe/H] within $\sim$0.1 dex).  These solar twin stars offer a unique opportunity to determine their compositions at a precision as good as 2\% \citep[see, e.g.,][]{Melendez2009, Ramirez2009, Bedell2014}.  We aim to combine this high-precision chemical analysis with the m s$^{-1}$-level detection capabilities of HARPS to investigate the star-planet connection at an unprecedented level of detail.

In this paper we present our first detection from the planet search program: a Jupiter twin around the solar twin \object{HIP11915}.  HIP11915 was previously established as a solar twin in \citet{Ramirez2014} based on its fundamental properties ($T_{\textrm{eff}}$ = 5760 $\pm$ 4 K, log $g$ = 4.46 $\pm$ 0.01, and [Fe/H] = -0.059 $\pm$ 0.004).  Its age, estimated from isochrone fitting to be 4.0 $\pm$ 0.6 Gyr, is also in keeping with the solar value \citep{Ramirez2014}.  HIP11915 is a bright ($m_V$ = 8.6) star with relatively low activity (log($R'_{HK}$) < -4.8), making it an excellent target for high-precision RV monitoring.  Based on a Keplerian signal in our HARPS RVs, we determine the planet candidate's orbital parameters.  We also investigate the influence of stellar activity on our RV measurements.

\section{Data}
All of the RV data used in this paper were obtained with the HARPS spectrograph on the ESO 3.6 m telescope at La Silla Observatory and were processed with the dedicated HARPS pipeline.  Forty-three data points spanning a time range from October 2009 to January 2015 came from our HARPS large program dedicated to solar twin stars (program ID 188.C-0265).  An additional 13 measurements were taken from the publicly available archive data and span a time range from October 2003 to November 2013 (program IDs 072.C-0488, 089.C-0732, 091.C-0034, 092.C-0721, and 183.C-0972).

For the 43 measurements taken through our program, the star was observed in high-accuracy mode with the ThAr lamp as a simultaneous reference source.  Each observation used a 900-second exposure time to minimize potential RV noise from the five-minute p-mode oscillations of solar-type stars.  To obtain a radial velocity measurement, we selected a binary mask based on the G2 spectral type for cross-correlation by the pipeline.  Owing to their generally high signal-to-noise ratios (median S/N over the spectral range on the order of 100), the pipeline-produced RV error estimates were generally below 1 m s$^{-1}$.  We expect the actual uncertainties to be somewhat higher owing to stellar noise and potential undiagnosed instrumental effects.  To account for these factors, we added the pipeline-estimated errors in quadrature with a baseline error level of 1 m s$^{-1}$, following \citet{Dumusque2011} and in accordance with the scatter in our own HARPS survey's quietest stars.

The 13 archive measurements have exposure times ranging from 120 to 900 seconds.  All were taken in high-accuracy mode without a source on the simultaneous reference fiber, so that even the low S/N measurements are not in danger of spectral contamination by the simultaneous reference source.  The data were processed by the pipeline and we modified the error estimates     as described above.  An additional 1.1 m s$^{-1}$ was added in quadrature to the errors on all exposures under 15 minutes, based on the expected p-mode signal for a solar twin \citep{Bazot2011}.

We analyzed all of the reduced HARPS spectra to measure the Ca II H\&K activity indices $S_{HK}$ and log($R'_{HK}$) at each observation epoch.  This was done according to the methods of \citet{Lovis2011}.  Errors on $S_{HK}$ were estimated from photon noise.

Complete RV and $S_{HK}$ data are available in Table \ref{tbl:data}, along with the HARPS pipeline-estimated activity indicators bisector inverse span (BIS) and full width at half maximum of the cross-correlation function (FWHM).

\section{Analysis and results}
\label{methods_rv}
The RMS of the RVs is 6.5 m s$^{-1}$, indicating the presence of a signal stronger than the predicted stellar jitter of 2.1 m s$^{-1}$ \citep{Wright2005}.  Visual inspection and a periodogram analysis of the data clearly show a long-period trend in the stellar radial velocities.  We ran a grid search across a range of periods from 0 to 10,000 days and found that the minimum chi-squared value lies near 3600 days.

We inspected the data for signs of stellar activity manifesting in the RVs.  A strong correlation between radial velocity and the activity index $S_{HK}$ is seen in the measurements from recent years, although there is no correlation in the archival data.  Similar RV-correlated behavior is seen for other activity indicators including the bisector inverse slope (BIS), FWHM, $V_{span}$, and the biGaussian fit differential \citep[as defined by][]{Queloz2001,Boisse2011,Figueira2013}.  Doing a weighted linear least-squares fit and removing the trend with any of these activity indicators from the RVs results in reducing the correlations while preserving the long-period planet signal with high significance in a periodogram analysis.  We additionally attempted the prescribed method of \citet{Dumusque2011} of fitting a long-period Keplerian  first to an activity diagnostic and then (with period and phase fixed) the RVs.  This method failed to remove the long-period power from the RVs regardless of whether $S_{HK}$, BIS, or FWHM was used as the diagnostic.  The failure of these methods to successfully ``clean'' the data of its long-period signal suggests that the signal originates from a true Doppler shift rather than stellar activity, although activity is also present in the data.

\subsection{Best model: Keplerian signal + $S_{HK}$ correlation}
\label{mainfit}

In the results presented here, we account for stellar activity by fitting a linear trend between RV and $S_{HK}$ simultaneously with the Keplerian signal.  This way the strength of the correlation between activity indicator and RV is free to change as the scatter induced by the Keplerian signal is removed.  We use $S_{HK}$ as the primary tracer of activity owing to the unreliability of the BIS diagnostic for slow-rotating stars and the demonstrated efficiency of $S_{HK}$ in removing activity cycle signals \citep{Boisse2011,Meunier2013}.  Other potential diagnostics and alternative model choices are explored in Section \ref{othermodels}.

We ran a Markov chain Monte Carlo (MCMC) analysis to fit the radial velocity data to a model consisting of three components: a Keplerian signal, a linear correlation term with the $S_{HK}$ values, and an offset term.  This gives seven free parameters: the Keplerian parameters ($P$, $K$, $e$, $\omega$ + $M_0$, $\omega$ - $M_0$), where $P$ is the orbital period, $K$ is the RV semi-amplitude, e is eccentricity, $\omega$ is the argument of periastron, and $M_0$ is the mean anomaly at reference date $t_0$ = 2457000.0; a slope parameter $\alpha$ which relates the activity-induced change in RV to the corresponding change in $S_{HK}$; and an offset parameter, $C$.  We  include an eighth free parameter, $\sigma_{J}$, representing the jitter that must be added in quadrature with the estimated RV error to make the measurement uncertainties consistent with the scatter in the residuals.  This jitter physically corresponds to unknown Gaussian error that may arise from stellar noise or undiagnosed instrumental variations.  When inputting the data, we subtracted a sigma-clipped median value from RV, $S_{HK}$, BIS, and FWHM for simplicity.  This means that the offset parameter C is partially set by the difference between the median RV of the data set and the systemic RV of the star, but also includes the RV offset in the linear fit with the median-subtracted $S_{HK}$ values (or any other correlations used in the model; see Section \ref{othermodels}).

\begin{figure}
\vspace{-5pt}
\begin{center}
\includegraphics[trim=95 15 20 50,clip,scale=0.37,angle=270]{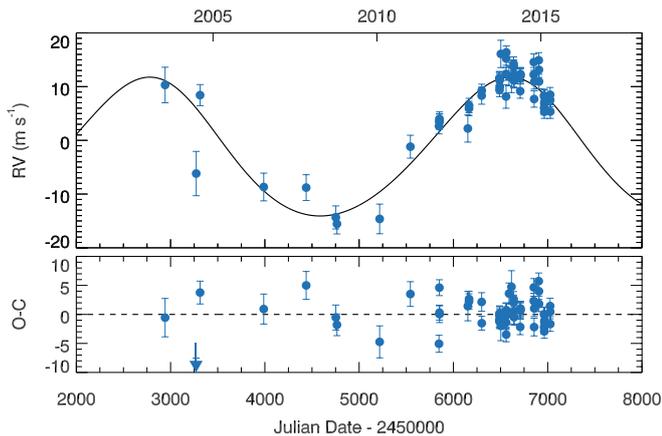}
\end{center}
\vspace{-10pt}
\caption{Best-fit orbital solution for HIP11915b.  Data points are the observed RVs with offset and $S_{HK}$ correlation terms subtracted.  One residual point is located outside of the plot range, as denoted by the arrow.}
\label{fig:orbit}
\end{figure}

\begin{figure*}
\includegraphics[scale=.42]{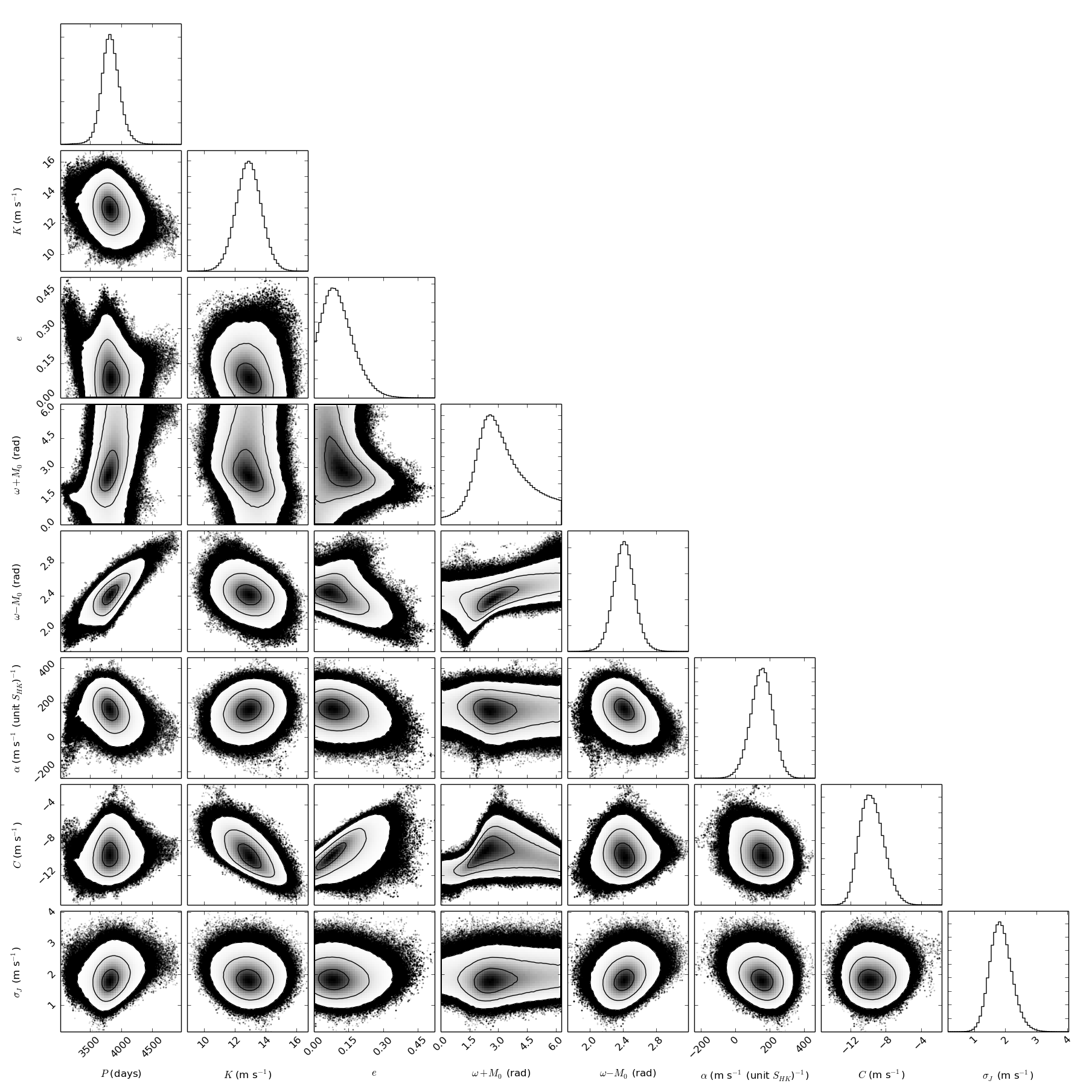}
\caption{Constraints on the fit to the radial velocities from the MCMC analysis.  Parameters are defined in Section \ref{methods_rv}.  Plot made using code from \citet{Foreman-Mackey:11020}.}
\label{fig:pairsplot}
\end{figure*}

All of the fit parameters were given uniform priors except for $P$ and $K$, which were sampled in log-space with a log-uniform prior.  We ran five parallel chains and used the requirement of a Gelman-Rubin test statistic below 1.01 for all parameters to avoid non-convergence \citep{Gelman1992}.  The best-fit orbital solution, plotted in Figure \ref{fig:orbit}, corresponds to an approximately Jupiter-mass planet on a 10-year orbit with eccentricity consistent with zero (see Table \ref{tbl:mcmc}).  As seen in the MCMC posterior distributions (Figure \ref{fig:pairsplot}), the chief source of uncertainty in the period measurement is in its degeneracy with eccentricity and the associated argument of periapse, parameters that are inevitably poorly constrained when only one orbit has been observed.  The best-fit jitter term is consistent with the predicted stellar jitter of 2.1 m s$^{-1}$ from the formula of \citet{Wright2005}.

\begin{figure}
\includegraphics[trim=30 0 10 0,clip,scale=0.32,angle=270]{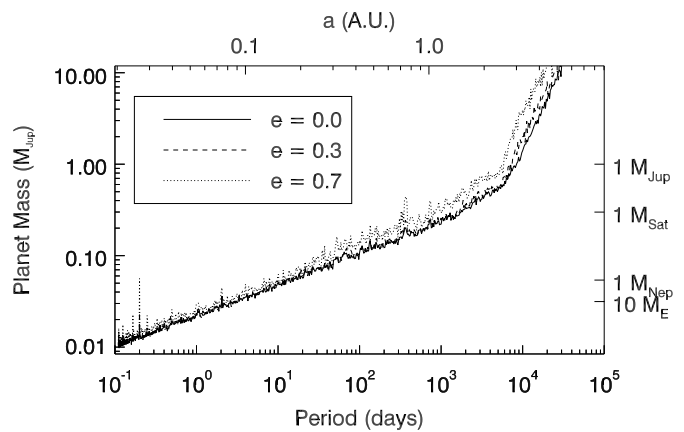}
\caption{Exclusion limits on potential planets in the RV data after subtraction of the Keplerian signal.  Any signal in the parameter space located above the lines plotted would have induced a larger scatter in the RV residuals with three-sigma significance if present.  Potential trends with activity indicators were not removed for this test to avoid underestimating the RV scatter.}
\label{fig:limits}
\end{figure}

After subtracting the best-fit planet signal, we searched for additional signals in the residuals with a generalized Lomb-Scargle periodogram analysis \citep{Zechmeister2009}.  No significant periodicities were identified in the residuals to the planet fit or in the residuals to the planet and the fitted $S_{HK}$ trend.  We place constraints on the presence of additional planets in the system using a test based on the bootstrapping and injection of a planet signal into the residuals.  For this test, we use the residuals to the planet fit only, and inflate the error bars (with jitter included) such that a flat-line fit to the residuals gives a reduced chi-squared of one.  We then run a set of trials in which the RV data are resampled with replacement at the times of observation, a planet signal with period and semi-amplitude fixed and the other orbital parameters randomized is injected, and the reduced chi-squared of a flat-line fit to these simulated data is measured.  If the chi-squared statistic has increased with three-sigma significance in 99 out of 100 trials at a given period and velocity semi-amplitude, we consider such a planet signal to be excluded based on its inconsistency with the scatter in the actual RVs.  We run this trial until upper limits on planet presence are found for all of the periods in a log-sampled grid.  The results rule out the presence of gas giants within a 1000-day orbital period, leaving open the possibility of inner terrestrial planets (Figure \ref{fig:limits}).

\begin{savenotes}
\begin{table}
\caption{Best-fit parameters and uncertainties for HIP11915b.}
\label{tbl:mcmc}
\centering 
\begin{tabular}{llcc} 
\hline    
\hline 
{Parameter}& & Value & Uncertainty  \\
\hline
$P$ &[days] & 3830 & 150 \\
$K$ &[m s$^{-1}$] & 12.9 & 0.8 \\
$e$ & & 0.10 & 0.07 \\
$\omega$ + $M_0$  &[rad] & 3.0 & 1.3 \\
$\omega$ - $M_0$  &[rad] & 2.4 & 0.1 \\
$\alpha$ & [m s$^{-1}$ (unit $S_{HK}$)$^{-1}$] & 160 & 60  \\
$C$ &[m s$^{-1}$] & -11.0 & 1.3 \\
$\sigma_{J}$& [m s$^{-1}$] & 1.8 & 0.4 \\
\hline   
$m_p$ sin($i$)  & [M$_\textrm{Jup}$]   & 0.99 & 0.06 \footnotemark[1] \\
$a$  & [AU]  & 4.8 & 0.1 \footnotemark[1] \\
\hline
$RMS$ &[m s$^{-1}$] & 2.9 & \\
\hline
\end{tabular}
\footnotetext[1]{Error estimates do not include uncertainties on the stellar parameters, which were taken from \citet{Ramirez2014}.}
\end{table}
\end{savenotes}

\subsection{Alternative model choices}
\label{othermodels}

It is not obvious from theory that the Keplerian +  $S_{HK}$ correlation model used above should be the best model choice.  Other activity indicators exist and have been used with success to remove activity trends from RVs in the past.  Chief among these are BIS and FWHM, both characteristics derived from the RV cross-correlation function (CCF) which measure the average spectral line asymmetry and/or broadening at the epoch of the observation \citep[see, e.g.,][]{Queloz2001,Queloz2009}. Either BIS or FWHM could potentially be used in place of $S_{HK}$.  Also, because these indicators arise from distortions to the rotationally broadened stellar absorption lines while $S_{HK}$ probes chromospheric emission, we could combine $S_{HK}$ with a CCF-derived measurement to simultaneously track two separate physical indicators of activity.

We considered several alternative models before settling on the Keplerian + $S_{HK}$ model as the best choice.  A summary of the models tried is given in Table \ref{tbl:models}.  In all cases, the fitting was done with an MCMC as described above.  The inclusion of a Keplerian signal is strongly justified, with the Keplerian +  $S_{HK}$ fit having a false alarm probability on the order of $10^{-15}$ compared to the best fitting activity model ($S_{HK}$ + FWHM).  While the lowest $\chi^2$ value is obtained for the fit to a Keplerian, $S_{HK}$, and FWHM, the addition of the FWHM term is not fully justified by the reduction in scatter around the fit.  An F-test gives a formal false alarm probability of 13\% for the Keplerian +  $S_{HK}$ + FWHM model compared to the simpler Keplerian +  $S_{HK}$ model.  The Keplerian +  $S_{HK}$ fit has a false alarm probability of 0.3\% when compared to the Keplerian only model.  Using BIS in place of $S_{HK}$ technically gives a slightly better fit, but we chose to use $S_{HK}$ in the main analysis because  of BIS's previously noted unreliability for slow rotators.  Regardless of the choice of activity indicator(s), all Keplerian fit parameters remain within the uncertainty intervals quoted in Table \ref{tbl:mcmc}.

\begin{table}
\caption{Models considered.}
\label{tbl:models}
\centering 
\begin{tabular}{lcc} 
\hline    
\hline 
{Name}& Degrees of Freedom & $\chi^2$ of Fit  \\
\hline
Keplerian only & 50 & 162.54 \\
Keplerian + $S_{HK}$ & 49 & 136.00 \\
Keplerian + FWHM & 49 & 145.32 \\
Keplerian + BIS & 49 & 133.96 \\
Keplerian +  $S_{HK}$ + FWHM & 48 & 129.50 \\
\hline
 $S_{HK}$ only & 54 & 897.78 \\
 BIS only & 54 & 704.27 \\
 FWHM only & 54 & 650.97 \\
 $S_{HK}$ + BIS & 53 & 719.28 \\
 $S_{HK}$ + FWHM & 53 & 639.76 \\
\hline
\end{tabular}
\end{table}

\section{Planet or activity cycle?}

The question of the role of stellar activity role in the observed radial velocities is of special importance given the proximity of the signal period to the 11-year solar magnetic cycle.  The signal size is also consistent with the theoretically predicted 10-11 m s$^{-1}$ signal for a solar activity cycle \citep{Meunier2013}.  However, an activity cycle should manifest not only in the observed RVs, but also in other diagnostics.  We expect a positive correlation in RV with the activity index $S_{HK}$ as well as with other diagnostics such as BIS and the FWHM of the cross-correlation function \citep{Lovis2011, Zechmeister2013}.  While such correlations are seen, they appear insufficient to fully explain the observed RVs.

For at least two other solar twins in our sample, similar Jupiter-like RV variations are seen; however, in these cases subtracting a linear fit with $S_{HK}$ entirely removes the long-period power from a periodogram and reduces the RMS of RVs to a level consistent with the expected stellar jitter.  In contrast, the RV signal inferred from the variation of $S_{HK}$ or any other activity tracer for HIP11915 is much smaller in size than the total RV signal, so that significant power remains at long periods in the periodogram. 

A large sample of stars with activity cycles have been studied with HARPS by \citet{Lovis2011}, who derived empirical predictions for the expected correlation between RV and other activity indictors with $R'_{HK}$ as a function of stellar temperature and metallicity.  The correlations seen for HIP11915 are generally lower than the predicted values: $C_{RV}$ = 30 $\pm$ 1 m s$^{-1}$ (unit $R'_{HK}$)$^{-1}$, $C_{FWHM}$ = 58 $\pm$ 2 m s$^{-1}$ (unit $R'_{HK}$)$^{-1}$, and $C_{BIS}$ = 22.4 $\pm$ 0.7 m s$^{-1}$ (unit $R'_{HK}$)$^{-1}$ from their Equations 9-12, while the $C$ values recovered from our data are 11, 45, and 16 respectively.  This disagreement suggests that the strength of activity seen in the data is insufficient to explain the RV signal.

\begin{figure*}
\centering
\includegraphics[scale=0.8]{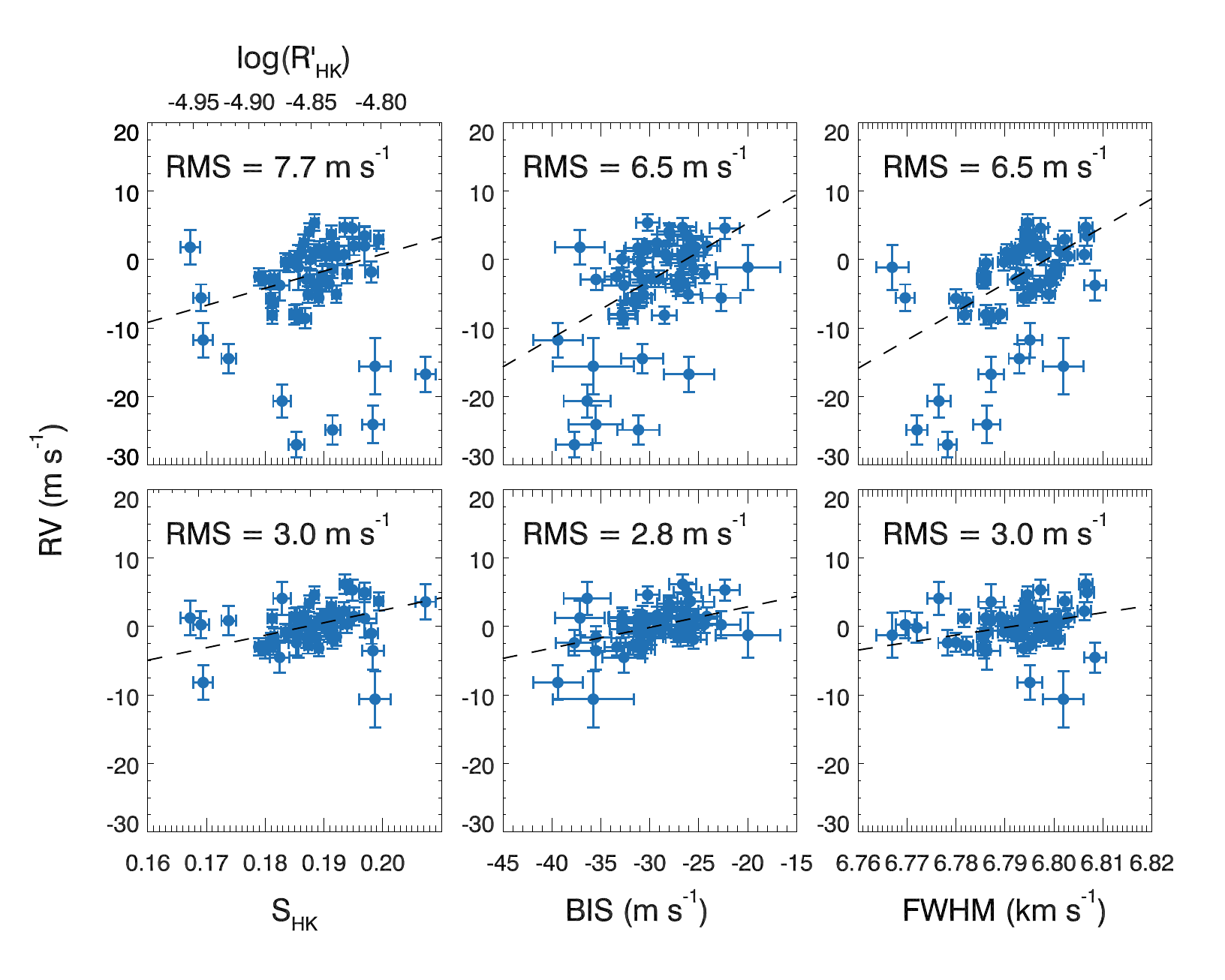}
\vspace{-10pt}
\caption{Correlations for the activity tracers $S_{HK}$, BIS, and FWHM with radial velocity.  The upper panels shows the observed RVs.  The lower panels show the residuals to the 3800-day signal fit (as presented in Section \ref{mainfit}).  Dashed lines are linear least-squares best fits to the data.}
\vspace{-10pt}
\label{fig:activity-rv}
\end{figure*}

An alternative interpretation is that we are observing activity overlaid with a planet signal in the radial velocities.  In this view, the correlation with activity indicator(s) in the RV residuals is independent of the Keplerian signal.  One piece of evidence in favor of this interpretation comes from inspecting the correlation plots before and after removal of the 3800-day Keplerian signal.  If this signal is truly unrelated to the stellar activity, it should add random scatter about the correlation.  Indeed, subtracting the Keplerian signal from the data reduces the scatter about the best-fit trend between RV and $S_{HK}$ from 7.7 to 3.0 m s$^{-1}$ and has a comparable effect on the correlations with other activity indices (Figure \ref{fig:activity-rv}). 

\begin{figure*}
\centering
\includegraphics[scale=0.9]{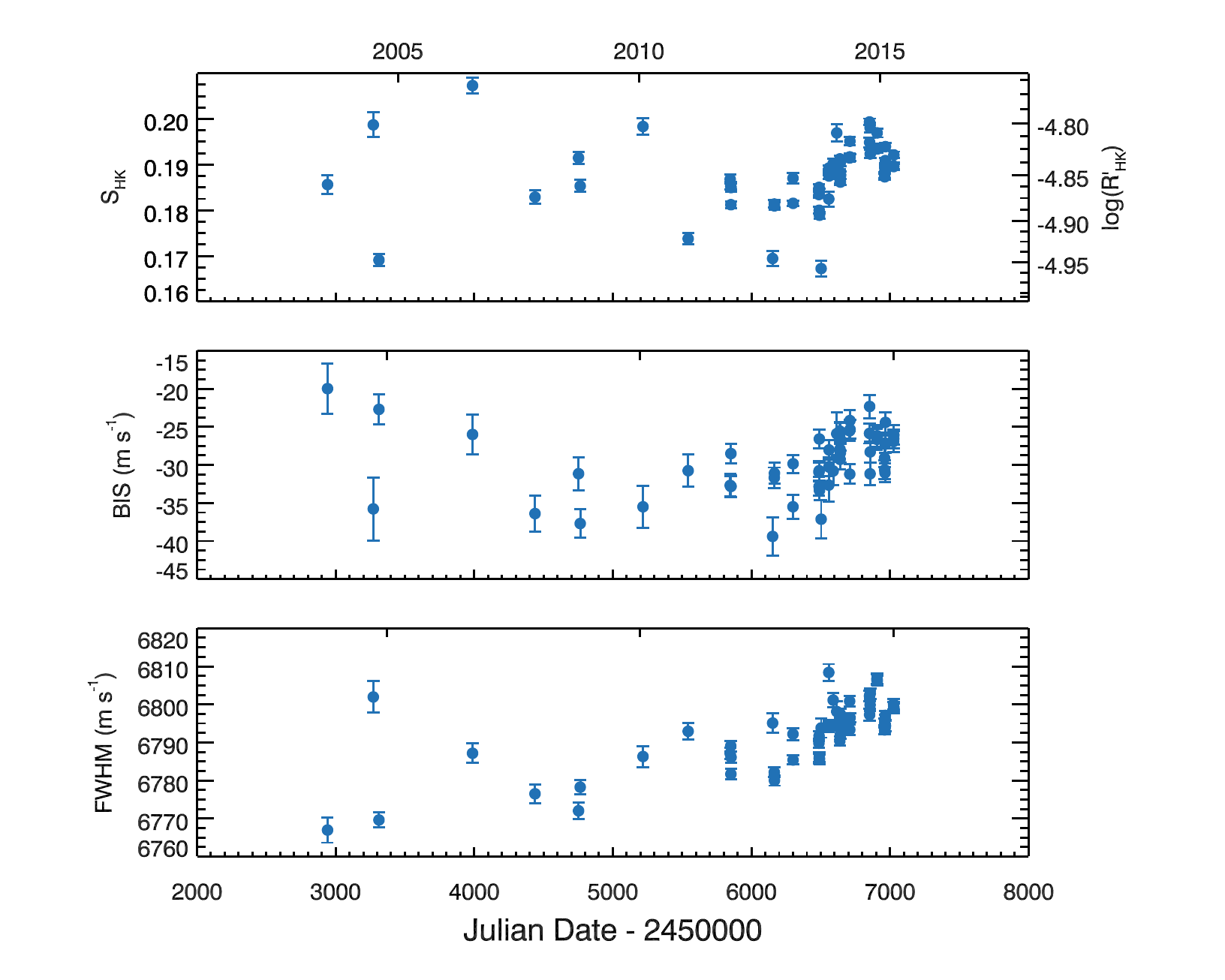}

\caption{Evolution of the activity indices $S_{HK}$, BIS, and FWHM with time.  Errors on BIS and FWHM are approximated as the errors on the RVs.}
\label{fig:activity-time}
\end{figure*}

Inspecting the behavior of $S_{HK}$ with time sheds some light on the evolution of HIP11915's activity level.  As depicted in Figure \ref{fig:activity-time}, the stellar activity level exhibited an incoherent scatter until mid-2013, at which point it increased from the equivalent of log($R'_{HK}$)$\sim$ -4.87 to $\sim$ -4.82 by mid-2014.  A concurrent rise in BIS and FWHM is also seen at this epoch.  This level of variability is consistent with the behavior of the Sun, which has a mean log($R'_{HK}$) of -4.94 and an amplitude of variation around 0.1 dex over a full activity cycle \citep{Hall2007}.  Given the recent increase in Ca emission, we could then expect a corresponding increase in short-duration activity phenomena such as spots and plages to manifest in the more recent RVs.  This would add power to the correlations with line asymmetry indicators. 

Ideally, we would be able to remove activity phenomena such as spots and plages by averaging our measurements over the timescale of the stellar rotation period, on the order of a month.  Unfortunately the sampling of the data is too uneven to average effectively over the rotation period.  We average instead over an entire observing season, which is a regime in which the time coverage should be distributed evenly enough to remove activity signals at periods shorter than a cycle.  The resulting data from this seasonal binning trace the 3800-day signal well, but show no statistically significant correlations with activity indicators.

In short, we have considered a variety of activity diagnostics and methods of cycle removal and the 3800-day signal remains robustly present in the data.  The coincidence in phase between the rise in $S_{HK}$ and the 3800-day signal maximum, however, calls for further monitoring of the star before we can definitively rule out the possibility of an activity cycle inducing the RV signal.

\section{Summary}

We have detected a Keplerian signal corresponding to a gas giant planet with a 10-year orbital period, consistent with a Jupiter twin, around the solar twin HIP11915.  While we cannot conclusively exclude an activity cycle interpretation, commonly employed activity tracers fail to account for the signal.  If the 3800-day signal is entirely induced by an activity cycle, the failure of various techniques to remove it bodes poorly for the prospect of correcting for such cycles in future long-period planet searches.

If this signal is truly planetary in origin, the HIP11915 system is a close analog to the solar system.  Our analysis shows the planet HIP11915b to be a close match to Jupiter both in mass and orbital period, and its host star is extraordinarily similar to the Sun.  Beyond having fundamental physical properties close to the Sun, initial spectroscopic analysis suggests that HIP11915 is also a solar twin in the sense that its detailed abundance pattern matches the solar pattern (Melendez et al. 2015, in prep).  The presence of a Jupiter twin and a solar-like composition both make HIP11915 an excellent prospect for future terrestrial planet searches.

\begin{acknowledgements}
We thank Jason Wright and Gaspare Lo Curto for helpful discussions.  We are also indebted to the many scientists and engineers whose past work on and observations with HARPS made such a long-period planet discovery possible.  M.B. is supported by the National Science Foundation (NSF) Graduate Research Fellowships Program (grant no. DGE-1144082).  J.M. would like to acknowledge support from FAPESP (2012/24392- 2) and CNPq (Bolsa de Produtividade). J.B. and M.B. acknowledge support for this work from the NSF (grant no. AST-1313119).  J.B. is additionally supported by the Alfred P. Sloan Foundation and the David and Lucile Packard Foundation.
\end{acknowledgements}

\bibliographystyle{aa} % style aa.bst 
\bibliography{HIP11915_ms} % your references Yourfile.bib

\clearpage

\clearpage

\begin{table}
\caption{HARPS measured radial velocities and activity indices for HIP11915.}
\label{tbl:data}
\centering 
\begin{tabular}{lccccccc} 
\hline    
\hline 
{Julian Date} & Exp. Time (s) & RV (km s$^{-1}$) & $\sigma_{RV}$ (km s$^{-1}$) & $S_{HK}$ & $\sigma_{S_{HK}}$ & BIS (km s$^{-1}$) & FWHM (km s$^{-1}$)  \\
\hline
 2452941.7219&116.5&     14.5766&  0.0030&  0.1856&  0.0020& -0.0200&      6.7670 \\
 2453271.8214&182.9&     14.5622&  0.0038&  0.1987&  0.0027& -0.0358&      6.8019 \\
 2453312.6877&900.0&     14.5722&  0.0017&  0.1691&  0.0013& -0.0227&      6.7696 \\
 2453987.7974&121.3&     14.5610&  0.0021&  0.2073&  0.0017& -0.0260&      6.7872 \\
 2454438.6089&134.1&     14.5571&  0.0019&  0.1829&  0.0015& -0.0364&      6.7765 \\
 2454752.7519&300.0&     14.5529&  0.0015&  0.1915&  0.0013& -0.0312&      6.7720 \\
 2454764.7361&900.0&     14.5507&  0.0016&  0.1853&  0.0013& -0.0377&      6.7782 \\
 2455217.5801&121.9&     14.5537&  0.0023&  0.1983&  0.0018& -0.0355&      6.7863 \\
 2455543.6010&500.0&     14.5633&  0.0015&  0.1738&  0.0012& -0.0308&      6.7929 \\
 2455846.8020&900.0&     14.5692&  0.0011&  0.1868&  0.0010& -0.0327&      6.7871 \\
 2455850.7947&900.0&     14.5698&  0.0009&  0.1854&  0.0009& -0.0328&      6.7891 \\
 2455851.7841&900.0&     14.5697&  0.0008&  0.1812&  0.0008& -0.0285&      6.7817 \\
 2455852.7779&900.0&     14.5697&  0.0010&  0.1850&  0.0009& -0.0328&      6.7861 \\
 2456152.8532&243.8&     14.5660&  0.0020&  0.1694&  0.0016& -0.0394&      6.7951 \\
 2456164.9273&900.0&     14.5716&  0.0009&  0.1814&  0.0008& -0.0317&      6.7822 \\
 2456165.9242&900.0&     14.5721&  0.0009&  0.1810&  0.0008& -0.0311&      6.7800 \\
 2456299.6151&900.0&     14.5749&  0.0012&  0.1870&  0.0011& -0.0355&      6.7922 \\
 2456300.5833&900.0&     14.5750&  0.0006&  0.1815&  0.0006& -0.0298&      6.7854 \\
 2456486.9106&900.0&     14.5774&  0.0007&  0.1834&  0.0007& -0.0309&      6.7898 \\
 2456487.8968&900.0&     14.5778&  0.0007&  0.1850&  0.0007& -0.0328&      6.7917 \\
 2456488.9425&900.0&     14.5749&  0.0008&  0.1800&  0.0007& -0.0308&      6.7857 \\
 2456489.9048&900.0&     14.5773&  0.0008&  0.1840&  0.0008& -0.0266&      6.7862 \\
 2456490.8972&900.0&     14.5753&  0.0007&  0.1789&  0.0007& -0.0334&      6.7856 \\
 2456502.9393&900.0&     14.5796&  0.0023&  0.1672&  0.0017& -0.0371&      6.7938 \\
 2456557.9085&900.0&     14.5740&  0.0020&  0.1825&  0.0016& -0.0326&      6.8084 \\
 2456558.8337&900.0&     14.5793&  0.0007&  0.1892&  0.0006& -0.0303&      6.7940 \\
 2456559.8500&900.0&     14.5819&  0.0007&  0.1875&  0.0007& -0.0280&      6.7945 \\
 2456560.8411&900.0&     14.5831&  0.0007&  0.1884&  0.0006& -0.0302&      6.7946 \\
 2456589.8019&400.0&     14.5790&  0.0010&  0.1902&  0.0010& -0.0308&      6.8011 \\
 2456615.6781&300.0&     14.5798&  0.0024&  0.1969&  0.0019& -0.0259&      6.7981 \\
 2456637.6474&900.0&     14.5789&  0.0007&  0.1884&  0.0007& -0.0256&      6.7904 \\
 2456638.6662&900.0&     14.5815&  0.0008&  0.1913&  0.0008& -0.0279&      6.7949 \\
 2456639.6356&900.0&     14.5802&  0.0008&  0.1866&  0.0007& -0.0293&      6.7955 \\
 2456640.6347&900.0&     14.5783&  0.0008&  0.1904&  0.0007& -0.0258&      6.7975 \\
 2456641.6456&900.0&     14.5786&  0.0007&  0.1862&  0.0007& -0.0284&      6.7921 \\
 2456643.6517&900.0&     14.5778&  0.0010&  0.1877&  0.0008& -0.0267&      6.7923 \\
 2456708.5010&900.0&     14.5764&  0.0009&  0.1915&  0.0009& -0.0255&      6.8008 \\
 2456709.5310&900.0&     14.5795&  0.0008&  0.1917&  0.0008& -0.0312&      6.7933 \\
 2456710.5234&900.0&     14.5798&  0.0009&  0.1951&  0.0009& -0.0242&      6.7964 \\
 2456711.5226&900.0&     14.5796&  0.0007&  0.1916&  0.0008& -0.0253&      6.7965 \\
 2456851.9220&900.0&     14.5807&  0.0009&  0.1993&  0.0007& -0.0259&      6.8022 \\
 2456852.8915&900.0&     14.5823&  0.0012&  0.1949&  0.0011& -0.0223&      6.7973 \\
 2456855.8977&900.0&     14.5760&  0.0011&  0.1981&  0.0011& -0.0312&      6.7999 \\
 2456856.8881&900.0&     14.5783&  0.0010&  0.1923&  0.0009& -0.0283&      6.8029 \\
 2456904.8198&900.0&     14.5825&  0.0010&  0.1936&  0.0009& -0.0266&      6.8065 \\
 2456906.7954&900.0&     14.5812&  0.0010&  0.1969&  0.0010& -0.0262&      6.8068 \\
 2456907.8079&900.0&     14.5785&  0.0009&  0.1934&  0.0009& -0.0264&      6.8062 \\
 2456960.7641&900.0&     14.5750&  0.0007&  0.1876&  0.0007& -0.0291&      6.7933 \\
 2456961.7200&900.0&     14.5726&  0.0007&  0.1873&  0.0006& -0.0306&      6.7951 \\
 2456963.7068&900.0&     14.5722&  0.0007&  0.1891&  0.0006& -0.0310&      6.7938 \\
 2456964.7840&900.0&     14.5737&  0.0007&  0.1887&  0.0007& -0.0271&      6.7961 \\
 2456965.6863&900.0&     14.5743&  0.0006&  0.1909&  0.0006& -0.0273&      6.7971 \\
 2456966.7461&900.0&     14.5757&  0.0009&  0.1939&  0.0008& -0.0244&      6.7943 \\
 2457025.5721&900.0&     14.5755&  0.0009&  0.1897&  0.0008& -0.0269&      6.8001 \\
 2457026.6043&900.0&     14.5745&  0.0007&  0.1896&  0.0006& -0.0266&      6.7993 \\
 2457027.5960&900.0&     14.5727&  0.0008&  0.1921&  0.0007& -0.0261&      6.7989 \\
 \hline
\end{tabular}
\end{table}

\end{document}